# STIMULATED RAMAN SCATTERING OF DYES UNDER RANDOM LASING IN POLYMERIC VESICULAR FILMS

V. P.Yashchuk[1], E.A.Tikhonov[2], O.A.Prygodiuk[1]
[1] Kyiv T. Shevchenko Nat. Univer., 03680, phys. depart., Glushkov ave. 2, b.1, Kyiv,
[2] Institute for Physics, Ukrainian Academy of Science, 02650, Nauky ave. 46, Kyiv

**Abstract.** The random lasing (RL) emission of dyed vesicular polymeric films in diffusive regime of light propagation was investigated under $(5 \div 8)^0$ K. It represents itself the coupling effect of stimulated emission and stimulated Raman scattering (SRS). Due to the effect a set of Raman lines reveal in RL spectrum. Distinguishing feature of the effect is the stimulated emission promotes the Raman lines revealing. Intensity of Raman lines have quadratic trend on pump intensity. Furthermore SRS in this condition has the more structured spectrum than spontaneous Raman one. It makes enable detection even the weak and close located spectral lines.

## INTRODUCTION

Recently the specific nonlinear phenomena — the coupling of stimulated Raman scattering (SRS) and stimulated emission (SE) into single process — was discovered in vesicular polymeric film with laser dyes Rhodamine 6G and pyrromethene 597 [1,2].

Under high pump intensity the stimulated emission of dyes in the multiply scattering medium dominates over the spontaneous one. It leads to the lasing-like emission formation named random lasing (RL). Spectrum of this emission depends on the scattering condition in a random media which is regulated by ratio between medium size $l_m$, mean scattering length $\bar{l_s}$ and emission wavelength λ.

Under diffusive regime of light propagation in random media ($\bar{l_m} \gg \bar{l_s} \gg \lambda$) the scattered waves are incoherent and they yield the total emission with continuous spectrum. Lasing-like emission in this regime is formed as ASE (amplified spontaneous emission) with continuous spectrum narrowed down to the minimal saturated value [3, 4]. But under certain pump intensity exceeding over RL threshold ASE spectrum of R6G radically changes. The spectrum widens and narrow spectral lines arise against the continuous spectra [1,2]. The lines spectral features indicate that they are conditioned by Raman scattering (RS). It gives the principal opportunity to study the Raman spectra from RL ones.

It is actually for lasing dyes because the powerful fluorescence often impedes weak signal of Raman scattering registration. Last years this problem was overcame by SERRS (surface enhanced resonance Raman spectroscopy) that increase RS cross section so the RS signal becomes comparable with fluorescence one. But this method needs very low concentration of the dye molecules so as to decrease the fluorescence. Therefore the RS emission is rather weak and may be registered with special sensitive technique only [5, 6]. Other possibility to obtain Raman spectra of lasing dyes exists under inverse Raman scattering [7]. But this technique is even more complicated because it requires two synchronized impulse sources of light: pumping monochromatic and probing with acceptable continuous spectrum in anti-Stokes region. At last single Raman lines corresponding to most high-Q modes may be observed under SRS but Raman spectra observation in this case is impossible.

RS obtaining in the case RL are free from these disadvantages. It doesn't require complicated technique of registration or specific light illumination. Instead one needs the dyed scattering media, laser pumping and single shot registration of radiation spectra. For the method development one have to study the conditions of RS spectra formation in joint ASE and SRS process and analyze the features of these spectra. These problems are the subject of this paper.

**EXPERIMENTAL**

**Samples and instrumentation.** We investigated RL spectra of the vesicular polymeric films dyed R6G, pyrromethenes 597 and 567. The films were fabricated from nitrogen-containing ternary copolymer. The ethanol solution of dye ($10^{-3} - 5*10^{-3}$M/l) was added to the solution of this copolymer and mixed. The mixture were deposited on the lavsan (dacron) substrate by coating. The film thickness after 3 sequential coating got equal $\cong$15mkm. Vesicles were generated after film polymerization by UV illumination and fixed by heating. Under this conditions nuclei of gaseous nitrogen are created and then associated in vesicles. Vesicle concentration depended on UV illumination doze. Under some sufficient doze vesicles got closely packed in the film plane. The vesicle mean diameter was about 1mkm. The microphoto (magnification $\sim$1000) of the dyed R6G film in reflected light is shown on the Fig.1.

Relative refractive index of the vesicles in used polymer was n$\cong$1.5. The high index refraction and close packing of vesicles provide multiple light scattering in sample plane, that is verified by significant ($\approx$10 nm) long-wave reabsorption shift of the fluorescence spectrum respectively to one in the neat film (Fig. 2). Similar result may be considered as evidence of high scattering efficiency in the film. Mean vesicle diameter ($\cong$1.2 mkm) corresponds to the diffusive condition $\overline{l_s} \gg \lambda$ in the film plane.

The emission spectra of the sample were single-shot registered by means diffraction spectrograph (grating 1200 mm$^{-1}$) and CCD camera. The total spectral resolution was 0.2 nm. Both spectra (lasing and fluorescence) were registered in the backward direction to pumping beam.

Image of the spectrum was digitized with custom-made computer program and displayed on the monitor. The program allows us to separate linear Raman component of spectrum from the continuous one conditioned by the random lasing.

Optical pumping of samples were executed by the second harmonic of Q-switched YAG:Nd$^{3+}$ ($\lambda$=532nm) laser. For checking if the spectral lines belonging to Raman scattering origin wavelength tunable phosphate glass: Nd$^{3+}$ laser was used. The laser pulse durations were $\tau \cong$15ns and 25ns respectively. Pumping beam was slightly focused on sample so the beam diameter at the side surface of the sample was $\cong$1mm. Pump intensity was varied with absorbing gray filters within the range $0.5*10^{-2} \div 0.5$ MW/mm$^2$.

To decrease Raman lines width and to shift RL spectrum to short wave direction so as to overlap it with Raman lines the experiments were performed at temperatures near to liquid helium (5-8)$^0$K. Other details of the experimental setup are shown on the Fig. 3.

**Results and discussion.** The radiation spectrum of dyed vesicular films under concentration R6G about $10^{-3}$M/l narrows to the saturation under the pump intensity $I_p$ up to the certain value $I_{th}$ and then becomes invariable under the further $I_p$ increasing (Fig. 4, curve 2). The $I_{th}$ value may be interpreted as the threshold pump intensity of a random lasing [8]. It corresponds to inflection point of the spectrum width dependence on pump intensity $\Delta\lambda(I_p)$. Due to lower positive feedback efficiency the spontaneous-to-lasing transition in RL is gradual (in contrast to abrupt one in conventional laser). The curve slope $\gamma=d(\Delta\lambda)/dI_p$ characterizes the rate of this transition. Both values $I_{th}$ and $\gamma$ depend on the dye concentration $C_d$: increasing results in $I_{th}$ lowering, $\gamma$ growth testifies the better lasing condition of RL.

RL with the more concentrated film has spectrum width under some times exceeding the threshold value (curve 2, Fig. 4). This spectrum widening is accompanying with the narrow lines appearance against the continuous background (Fig. 5). Their intensities increase with pump growth and its wavelengths are invariable and highly reproducible from shot to shot. Less dye concentrated film does not reveal this phenomenon: its spectrum is invariable under even more substantial exceeding of RL threshold (curves 1, Fig. 4).

The absolute reproducibility of the spectral lines under these conditions point out the line structure is not conditioned by film interference of scattered waves but is inherent to the investigated medium itself. The spectral lines in RL spectrum shift under wavelength variation of pump radiation in strong accordance with its values [2]. Wavelength of the lines in RL spectrum well coincide with ones of Raman spectrum obtained by the SERRS method [5, 6] and inverse Raman scattering [7]. This data proves explicitly that Raman scattering of dye molecules is source of the lines in the RL spectra.

The number of Raman lines revealed in RL spectrum depends on the spectrum width. The maximal RL spectrum width corresponds the maximal number of Raman lines that revealed under the higher R6G concentration. Corresponding RL spectrum and Raman spectrum separated from the first one are displayed on the Fig. 6a.

The answer why the Raman scattering occurs for dye molecules, whose concentration is much lower than polymer matrix molecules one can obtain from random lasing dependence on dye concentration (Fig. 6b). RL spectrum of R6G shifts to longwave direction with $C_d$ grows because of enhanced reabsorption in overlapping region of absorption and fluorescence spectra. One can observe that Raman lines reveal itself only if they appear in range of the RL spectrum. Background of RL spectrum corresponds to ASE and it reproduces in some approximation of the dye gain contour. Therefore the background shift reflects the same shift of the dye gain and ASE contours. In this connection Raman lines are revealed in RL only when these lines dispose in the gain (ASE) contour. In this manner ASE gain "brightens" Raman lines disposed in its range and makes them apparent.

That is evident from above Raman lines in RL spectrum appear when the dyed film are irradiated by two light flows simultenesly: pumping and random lasing flows (ASE). Therefore the Raman line intensity $I_s$ (relatively continuous background of RL) is proportional to the product of random lasing $I_{RL}$ and pumping $I_p$ intensities. Because magnitude $I_{RL}$ is proportional to the pumping power, the Raman lines intensity $I_s$ is depended in quadratic power on $I_p$. It is illustrated clearly by Raman lines in the RL spectrum of pyrromethene 597 where one of the lines (at λ=569.5 nm) dominated over other (Fig.7a) and due to this domination $I_s(I_p)$ dependence may be detected more accurately.

The combined action of the two mentioned radiations can be considered as specific display of active Raman scattering spectroscopy (it is more known as coherent anti-Stokes scattering CARS, because difference of their frequencies $\omega_p - \omega_s$ resonates with molecular vibration of frequency $\omega_m$ and swings them [9]. In the conventional method the samples are irradiated by bichromatic pump formed by two external monochromatic radiations: base pump with the fixed frequency $\omega_p$ an additional one with tunable frequency $\omega'$. The Stokes scattering of the third (probing) monochromatic emission of frequency $\omega$ is investigated in dependence on $\Delta\omega=\omega_p-\omega'$. This process may be conditionally depicted as $\omega_s = \omega - (\omega_p - \omega')$. The intensity of the scattered probing radiation depends on $\Delta\omega$ resonantly due to third order nonlinear susceptibility $\chi^{(3)} \sim (\omega_i^2 - \Delta\omega^2 + 2i\Gamma\Delta\omega)^{-1}$ and reach the maxima under condition $\Delta\omega=\omega_m$. In such way all active Raman frequencies are detected by tuning of the additional pump frequency $\omega'$.

In the strongly scattering active medium the additional component of bichromatic pumping is created inside the sample in the separate process – stimulated emission (random lasing) of dye molecules. RL radiation is developed in wide spectral range and serves as additional pumping at all Raman frequencies simultaneously. Owing to these peculiarities there isn't necessity of external bichromatic pump. Only the single monochromatic pumping at frequency ω is enough – second one arises in the oscillating sample itself. There isn't necessity of the special probing radiation also because its role plays the RL radiation at Raman frequencies $\omega_s$. That is why in the strongly scattering medium there is the degenerated case of active spectroscopy of Raman scattering in which external (base) pumping serves as probing radiation as well and RL radiation at Raman frequencies serves as additional pump: $\omega_s = \omega'$. Then this case may be conditionally depicted as $\omega_s = \omega_p - (\omega_p - \omega_s)$. Due to the continuous spectrum of RL in diffusive regime all Raman frequencies which occurred in its contour may amplify and reveal in the total radiation spectrum. From this point of view the RL radiation (conditioned by ASE) is the additional pumping with broad spectrum and then there isn't necessity to tune its frequency. The SRS of dye molecules is resonant because the pump wavelength coincides with dye absorption band. Therefore SRS probability of dye molecules is much higher than the matrix one. For this reason only dye Raman lines can be observed in RL emission.

*By this means dye RL emission in the strong scattering media is conditioned by two processes: ASE and amplified Raman scattering at all Raman frequencies occurring in ASE spectrum.* Correspondingly the RL spectrum represents itself the sum of the ASE (solid background) and Raman scattering (lines) component. With this understanding the radiation intensity $I_L$ at RS frequency $\omega_s$ is the sum of RL contribution $I_{RL}$ (conditioned ASE) and Raman scattering one $I_S$ (Fig.8):

$$I_L = I_S + I_{RL}. \tag{1}$$

$I_S$ and $I_{RL}$ increments per unit length can be described by the set of coupled differential equations [10, 11]:

$$\frac{dI_S}{dz} = g \cdot \chi^{(3)}(\omega_S) I_p I_L, \tag{2}$$

$$\frac{dI_{RL}}{dz} = (\sigma(\omega_S) \cdot \Delta(I_p) - \alpha) I_L, \tag{3}$$

where $\chi^{(3)}(\omega_s)$ is the value of cubic nonlinear susceptibility at Raman frequency $\omega_s$ (then $\Delta\omega = \omega_p - \omega_s$); g — proportional coefficient depending from Q- factor of the molecule vibrations; $\sigma(\omega_s)$ — transition cross-section of stimulated emission between the dye lasing levels; Δ — sample-averaged inverse population density; α - loss coefficient conditioned by the absorption and sample overrunning of the radiation.

Solution of system (3) yields the following expression of the lasing intensity:

$$I_L = I_{L,0}(\omega_s, I_p) \exp\left(\left(g \cdot \chi^{(3)}(\omega_s) \cdot I_p + \sigma(\omega_s) \cdot \Delta(I_p) - \alpha\right) \cdot l\right), \quad (4)$$

where $I_{L,0}$ - sample-averaged fluorescence intensity. Expression (4) is valid for any frequencies (not only for Raman one) due to frequency-dependent parameters $\chi^{(3)}$ and $\sigma$. According to (4) RL spectrum is formed by coupling of stimulated emission and stimulated Raman scattering that yield quasi-linear spectral dependence of the gain coefficient:

$$k(\omega) = g \cdot \chi^{(3)}(\omega) \cdot I_p + \sigma(\omega) \cdot \Delta(I_p). \quad (5)$$

Owing to this dependence the quasi-linear contour of RL spectrum is formed as the result of amplification of continuous fluorescence spectrum $I_{L,0}(\omega)$. When Raman line intensity is small ($I_S << I_{RL}$) the expression for their intensity can be derived from (4) and written as:

$$I_S = g \chi^{(3)} I_{RL} I_p l \sim I_p^2, \quad (6)$$

where $I_{RL} \sim I_p$ [3, 8, 11]. The expression (6) agrees with experimental dependence (Fig.7b) and therefore proves the validity of the consideration.

The expression (4) can be presented also in the following form:

$$I_{RL}(\omega) = I_{lum}(\omega) \cdot G_{se}(\omega) \cdot G_{ss}(\omega) \quad (7)$$

where $I_{RL}(\omega)$ and $I_{SRS}(\omega)$ are the spectrum contours of RL (conditioned by ASE) and SRS correspondingly which are represented by the factors of the expression (4):

$$I_{RL}(\omega) = I_{L,0}(\omega_s, I_p) \exp\left(\left(\sigma(\omega_s) \cdot \Delta(I_p) - \alpha\right) \cdot l\right), \quad (8)$$

$$I_{SRS} = \exp\left(g \cdot \chi^{(3)}(\omega_s) \cdot I_p \cdot l\right). \quad (9)$$

The expression (7) allows to write the spectral dependence of cubic susceptibility $\chi^{(3)}(\omega)$ as:

$$\chi^{(3)}(\omega) \sim \ln I_{SRS}(\omega) = \ln\left(\frac{I_L(\omega)}{I_{RL}(\omega)}\right) \quad (10)$$

Expression (10) presents SRS spectrum of the dye molecule in arbitrary units. For practical use (10) one needs to divide the total RL spectrum $I_L(\omega)$ (containing ASE contour and SRS spectra) by the continuous background of it (corresponding only to ASE). To mark the way of Raman spectrum obtaining we designate it SRS-RL. Results of computation for R6G in comparison with SERRS spectra taken from two independent sources [5, 6] is shown on the Fig.9a. For convenience maxima of the spectral lines are marked with dropped lines. One can see the computed SRS-RL spectrum agrees very well with SERRS ones. Majority of frequencies of these spectra coincide each other; its relative intensity very close too. But the lines in SRS-RL spectra are sharper and better separated. Due to that several additional lines are resolved in the SRS-RL spectrum but it isn't resolved in SERRS ones (such as lines λ = 565.3, 566.1; 570.9, 571.9; 575.9, 576.9, 577.6 nm). Reality of these resolved components is confirmed by the IR absorption spectrum where they are trusty detected [12].

Reduced results are more obvious from Fig.9b, c, where deconvolution on Gauss spectral forms are shown. The SERRS spectrum is decomposed on four Gauss forms much better than SRS-RL one. One can see that it stipulated by more structured SRS-RL spectrum as compared with SERRS one. Lateral structure of the four lines in SERRS spectral fragment is barely noticeable but in the SRS-RL one this structure is trusty registered. Due to the better structured SRS-RL spectrum it can be decomposed on more Gauss forms to reach the same quality of fitting as SERRS spectrum - nine Gauss components at least. SERRS spectrum is not exemplar for comparison because of inhomogeneous broadening but it is the best Raman spectrum of R6G (except the inverse Raman scattering) available now: conventional resonance Raman spectra are not available because intensive dye fluorescence. Better resolution of the lines in SRS-RL spectrum evidence about perspective of this method for Raman spectra studying.

The probable reason of greater structuring of SRS-RL spectrum is the coherent nature of both emission mechanisms formed the RL spectrum: stimulated emission and stimulated Raman scattering. Due to that radiation of different process may interferes each other and makes dipper the space between lines like the interference of resonant and non-resonant components of the cubic nonlinear susceptibility of SRS in CARS [9]. Multiple scattering on the vesicles destroy partially the coherence of the waves but it might keep it at space between two consequence scattering. Raman amplifying as other reason of the structuring might be excluded because it doesn't influence the line contour in representation (10). In consideration of this circumstance the electric field strength of the ASE and Raman scattering waves should be sum in (1) instead of intensity. This approach is expected should be able more suitable to describe the SRS-RL spectrum.

**SUMMARY**

RL of dyes in diffusive propagation represents itself coupling of ASE and SRS. Due to this coupling a set of the active in Raman scattering lines may reveal in RL spectrum. The expression for Raman spectra calculation from SRS – RL spectra was presented. By quantitative comparison of observed Raman spectrum and one registered with other methods was shown they are the same. Therefore presented method can be considered as new good way of Raman spectrum study.

It is important that the intensive fluorescence doesn't impede the Raman lines registration but promote that because Raman line intensity is proportional to ASE. Appreciated advantage of the method is the greater structuring of the SRS-RL spectrum conditioned by coherent nature of the both coupling process. It makes enable detection the weak and close on freuquency lines as it is illustrate by the fig.9 and confirms by IR absorption spectrum.

The peculiarities of the studied phenomenon are rather important for the lasing dyes spectroscopy because they make possible to study Raman spectrum by means of excitation of RL spectra without the special technique. Helium temperature promotes the Raman lines observation but it is not obligatory. For application of this method the strong scattering dyed media have to be prepared only. In the presented work we used the vesicular films which are very effective scattering system but the concentrated suspensions of dielectric particles (in polymer) are suited also.

## TITLES OF THE PAPER FIGURES

Fig.1. Microscopic image of the dyed vesicular film in reflected light.

Fig.2. R6G fluorescence spectrum (2) of vesicular in compare to one of homogeneous film (1) and RL ($2^/$) spectrum.

Fig.3. Experimental setup scheme.

Fig.4. Dependence of RL spectrum width on pumping for the various R6G concentrations: $10^{-3}$M/l (1) and $3*10^{-3}$ M/l (2).

Fig.5. Structures in R6G spectra (1,2) and its 2D-images (1´, 2´) for higher concentrated ($3*10^{-3}$M/l) samples under pumping indicated as $I_{p1}$ (1, 1´) and $I_{p2}$ (2, 2´) on the Fig. 4.

Fig.6. a) Spectra of RL R6G for concentration $5*10^{-3}$M/l (1) and SRS (2) separated from the first one; b) Shift of RL spectra and gain contour (background) in compliance with R6G concentration increase from 2mM/l (1) to 3mM/l (2) and revealed sections of dye Raman spectrum: $1^/$ — R6G concentration is 2mM/l, $2^/$ — 3mM/l , $3^/$ — 5mM/l,

Fig.7. a) Experimental RL (1) and calculated Raman (2) spectra of pyrromethene 597; b) dependence of the most insensitive Raman lines λ=569.5 nm from pumping.

Fig.8. Components of coupled emission in RL: stimulated emission (ASE)- $I_{RL}$ and SRS- $I_s$

Fig.9. Comparison of SERRS (1 — [6], 2 — [5]) and SRS (3) spectra (a); deconvolution of SERRS ((b) — [6]) and SRS (c) spectrum fragments on Gauss contours.

**FIGURES TO PAPER**

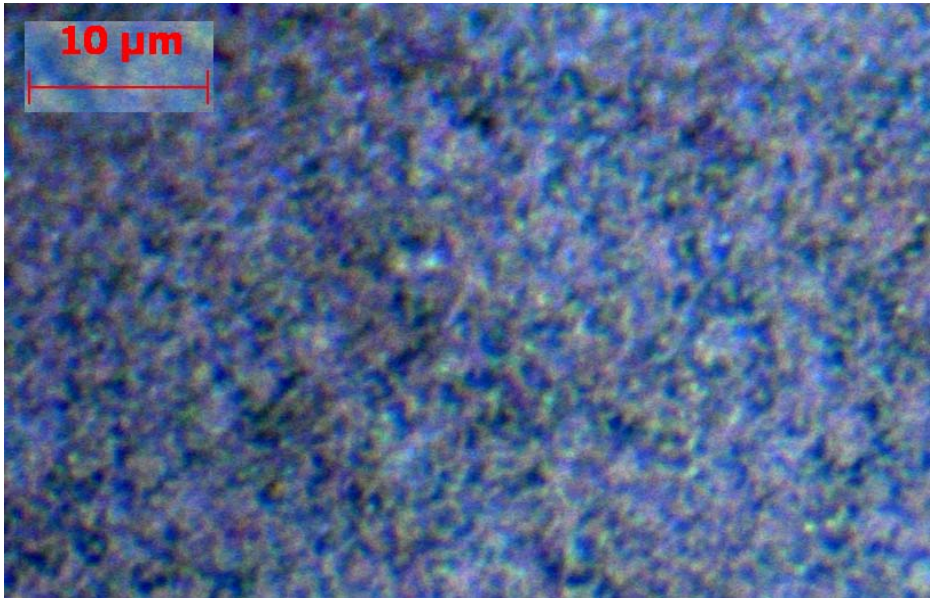

Fig.1.

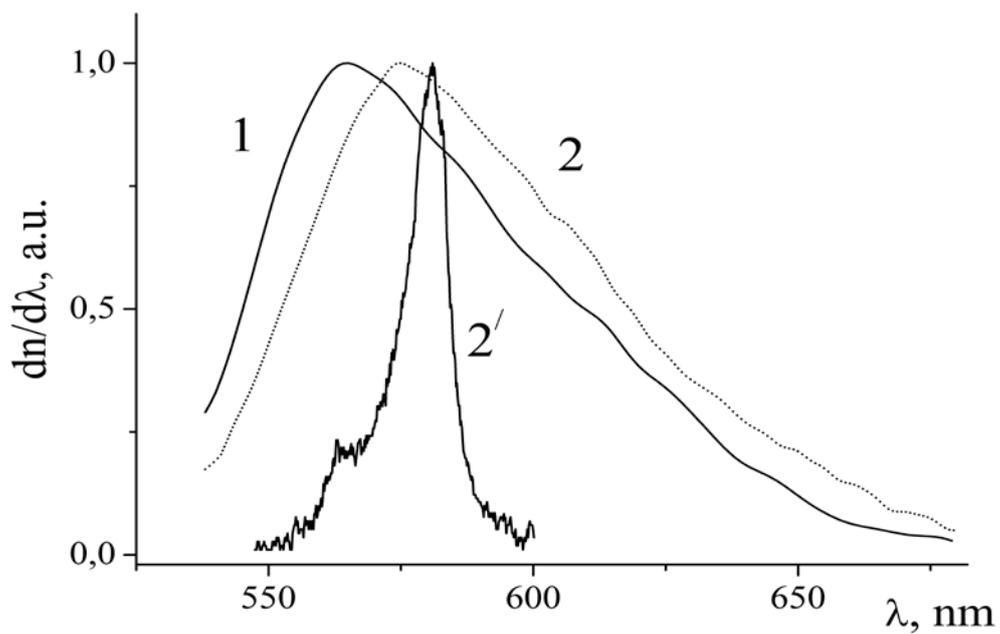

33

Fig.2.

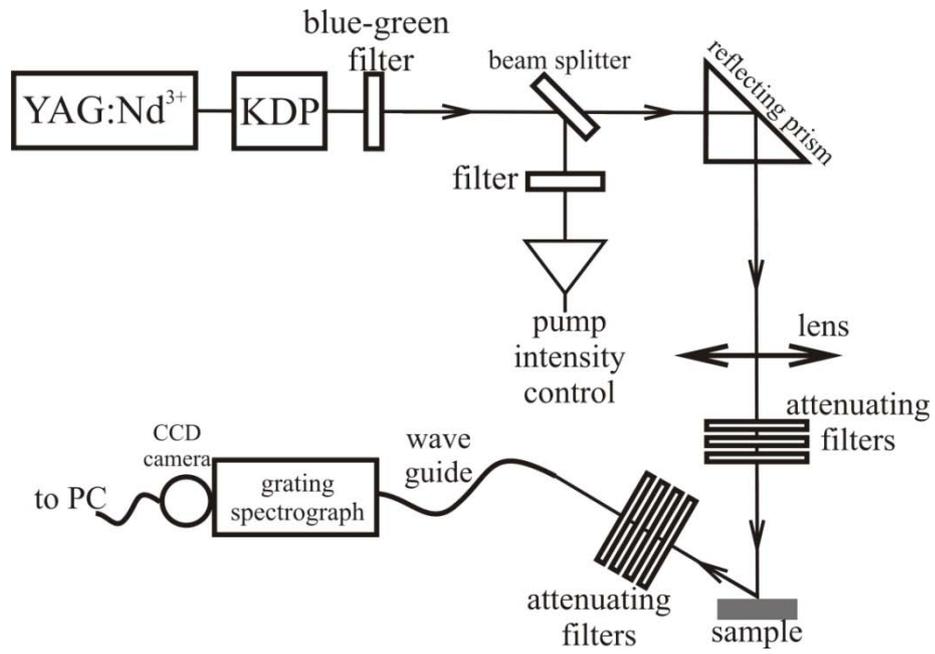

Fig.3.

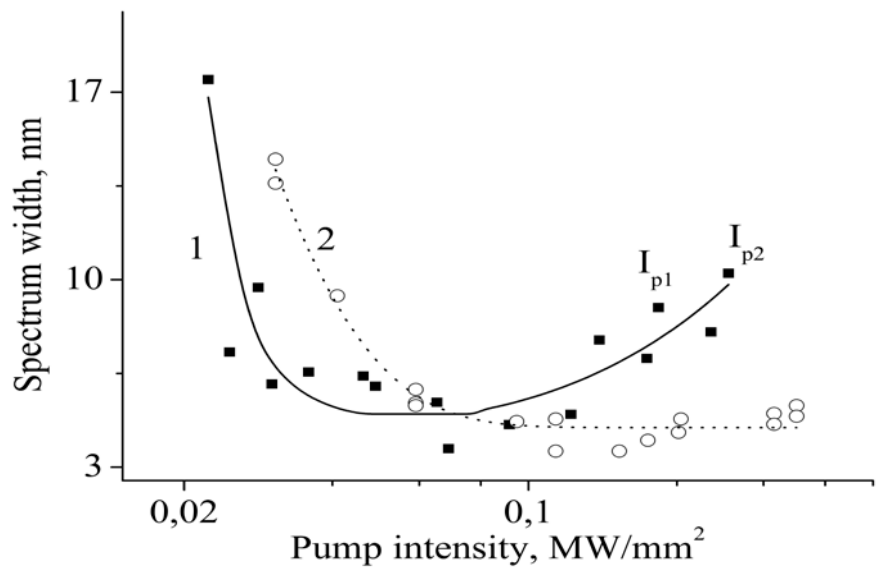

Fig.4.

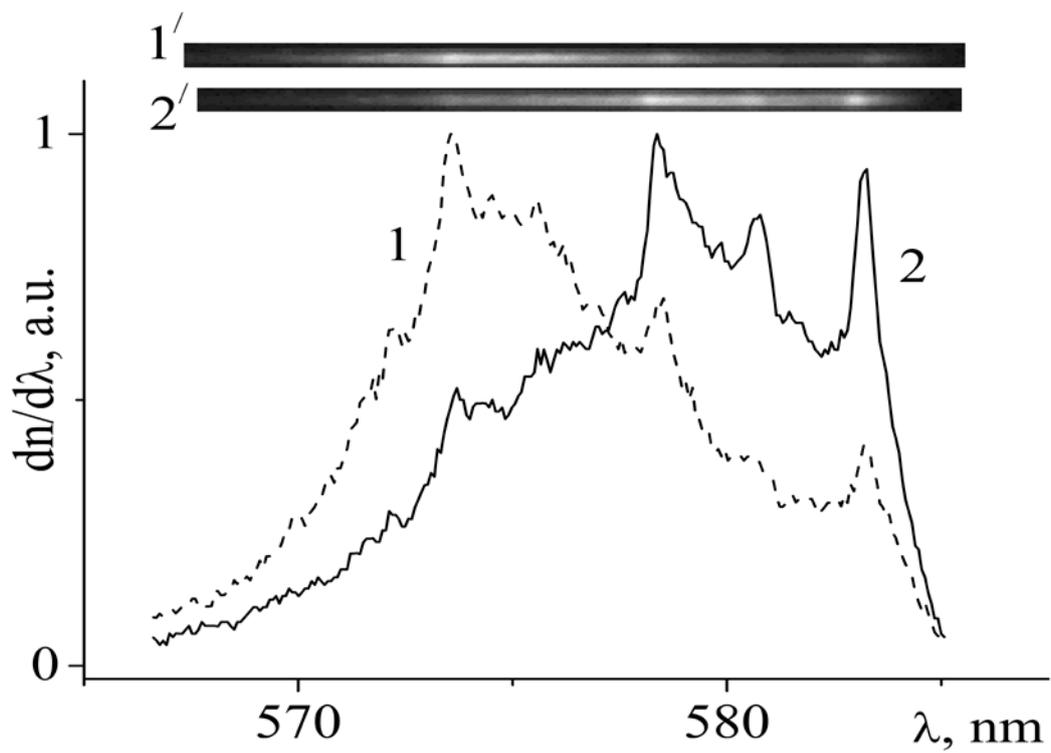

Fig.5.

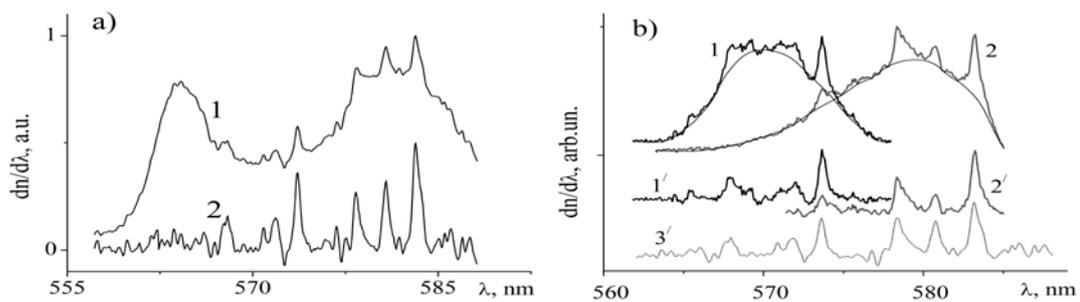

Fig.6.

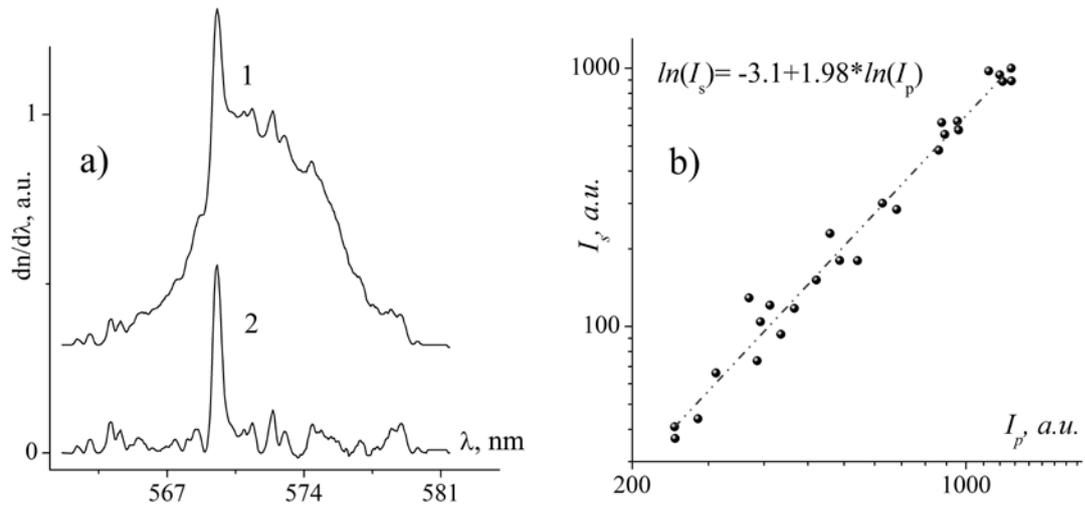

Fig.7.

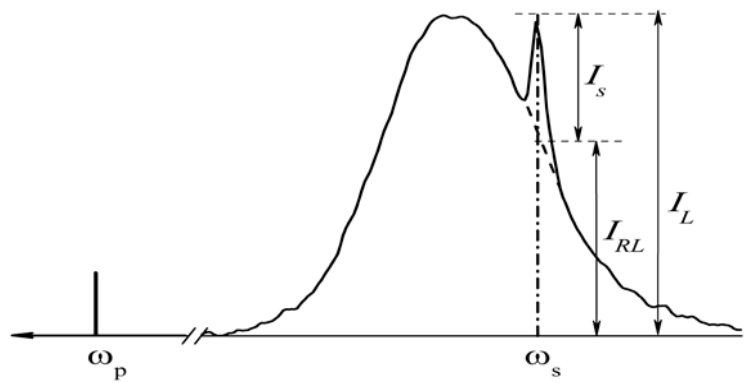

Fig.8.

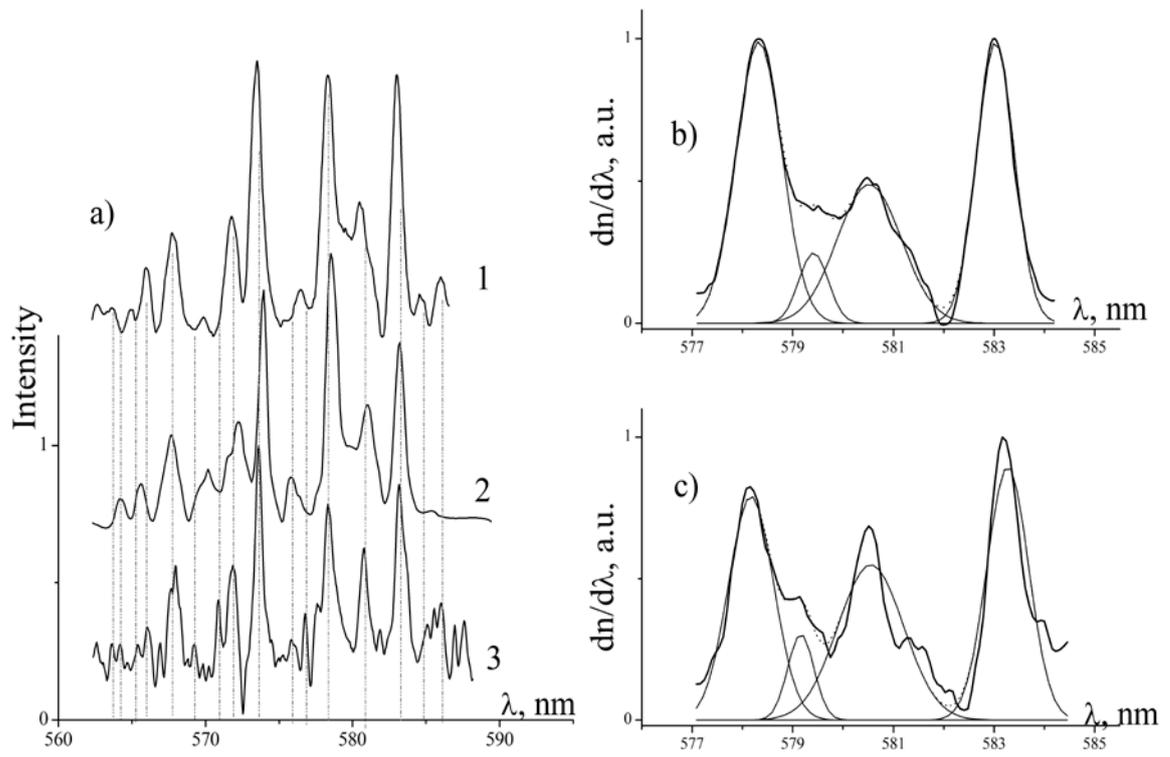

Fig.9.